# The Potential of Social Media Analytics for Improving Social Media Communication of Emergency Agencies

**Milad Mirbabaie**
Department of Computer Science and Applied Cognitive Science
University of Duisburg-Essen
Duisburg, Germany
Email: milad.mirbabaie@uni-due.de

**Jennifer Fromm**
Department of Computer Science and Applied Cognitive Science
University of Duisburg-Essen
Duisburg, Germany
Email: jennifer.fromm@uni-due.de

**Simone Löppenberg**
Department of Computer Science and Applied Cognitive Science
University of Duisburg-Essen
Duisburg, Germany
Email: simone.loeppenberg@stud.uni-due.de

**Sophie Meinig**
Department of Computer Science and Applied Cognitive Science
University of Duisburg-Essen
Duisburg, Germany
Email: sophie.meinig@stud.uni-due.de

**Matthias Reuße**
Department of Computer Science and Applied Cognitive Science
University of Duisburg-Essen
Duisburg, Germany
Email: matthias.reusse@stud.uni-due.de

## Abstract

A growing number of people use social media to seek information or coordinate relief activities in times of crisis. Thus, social media is increasingly deployed by emergency agencies as well to reach more people in crisis situations. However, the large amount of available data on social media could also be used by emergency agencies to understand how they are perceived by the public and to improve their communication. In this study, we examined the Twitter communication about the German emergency agency "Johanniter-Unfall-Hilfe" by conducting a frequency, sentiment, social network and content analysis. The results reveal that a right-wing political cluster politically instrumentalised an incident related to this agency. Furthermore, some individuals used social media to express criticism. It can be concluded that the use of social media analytics in the daily routine of emergency management professionals can be beneficial for improving their social media communication strategy.

# 1   Introduction

A growing number of people turn to social media to interact with family members and friends, share entertaining videos, discuss opinions with others and seek information about brands, products or events (Whiting and Williams 2013). The extensive usage of social media is leading to an increasing amount of available information which could be of interest for organisations. Private sector organisations, for example, already use social media monitoring tools to gain insights into how their brand is perceived by social media users (Bekmamedova and Shanks 2014). These insights are valuable for companies, as they can adjust their social media communication strategy accordingly. Furthermore, social media data advantage the understanding of customer needs and may drive effective advertising and marketing strategies (Bekmamedova and Shanks 2014).

While a lot of companies have developed a social media strategy to establish a bidirectional communication channel with (potential) customers and gain insights based on social media analytics, non-private organisations seem to be more hesitant to implement social media into their operations. At least in Germany, emergency agencies belong to this kind of organisations; however, they are currently starting to experiment with social media. At the present time, emergency agencies in Germany use social media primarily as another broadcasting channel to disseminate warnings or behavioural advice in times of crisis and give insights into their work practices (Eismann et al. 2016). Previous studies have revealed that emergency management professionals are hesitant to use social media analytics to gain additional insights due to a lack of guidelines and training (Stieglitz, Mirbabaie, Fromm, et al. 2018). Nevertheless, they are convinced that social media communication and analytics will be an important part of their future work routine (Reuter et al. 2015).

Therefore, this research seeks to demonstrate the potential of social media analytics for improving the social media communication of emergency agencies. As organisations assisting people in emergencies and crisis situations, it is of high importance for them to maintain a positive and trustful image. The use of social media analytics can provide emergency agencies with insights into how they are perceived by the public. Hence, we examine the following research question:

**RQ: Which insights into the communication about emergency agencies can be gained through social media analytics?**

In this research, we use different methods of social media analytics (frequency analysis, sentiment analysis, social network analysis and content analysis) to analyse the Twitter communication about the German emergency agency "Johanniter-Unfall-Hilfe (JUH)". This emergency agency is a voluntary humanitarian organisation focusing on training, emergency medical service, disaster relief, social care and international help (Johanniter-Unfall-Hilfe 2018). We decided to analyse the communication on Twitter because the microblogging platform has the advantage of a detailed Application Programming Interface (API), simplifying the access to data (Twitter 2018). Based on the results of the analysis, we derive implications for the use of social media analytics to improve the social media communication of emergency agencies. In the next section, we summarise related work on social media communication and analytics focusing on possible usage of emergency agencies. Then, we describe our research methods and the corresponding results. Afterwards, the implications of these results for the social media communication strategies of emergency agencies are discussed. At the end, the limitations of this study and avenues for further research are presented.

# 2   Background

## 2.1   Emergency Agencies and Social Media Communication

Social media platforms are increasingly used by citizens to make sense of crisis situations, seek information or coordinate relief activities (Mirbabaie and Zapatka 2017; Stieglitz, Bunker, et al. 2017; Stieglitz, Mirbabaie, and Fromm 2018). Meanwhile, social media platforms themselves offer functionalities (e.g. Facebook Safety Check) which enable users to indicate whether they are safe in times of crisis (Takahashi et al. 2015). As social media platforms allow to reach a larger proportion of the population in crisis situations, these platforms are nowadays used by media organisations and emergency agencies as well (Bruns and Burgess 2014). Interviews with emergency management professionals have revealed that they perceive social media as important for presenting their organisation, recruiting volunteers and establishing relationships with other emergency agencies or media organisations (Reuter et al. 2016). As emergency agencies strongly depend on recruiting new generations of volunteers, the relevance of social media for successful self-presentation and establishing relationships to the public increases. In practice, however, the social media usage of emergency agencies is often limited to broadcasting information to the public via Facebook or Twitter, thus the potential of social media for two-way communication is not fully exploited (Eismann et al. 2016). Emergency agencies currently use these platforms to provide the public with short situational updates, warnings, advice and guidance on how to prevent crises or how to behave during crises (Fosso Wamba and Edwards 2014). After a crisis has been resolved, emergency agencies often publish a final report to reduce uncertainties about what has happened (Reuter et al. 2016). In addition, they make attempts to correct misinformation and counter rumors on social media (Plotnick et al. 2015). An analysis of the tweets of 100 non-profit organizations showed that emergency agencies use Twitter primarily to share information about their organisation and activities but, to a lesser extent, also for fostering relationships to media organisations and the public or calling for actions (Lovejoy and Saxton 2012). Other studies examining the Twitter communication of the Queensland Police Service or the Boston Police Department led to similar results (Ehnis and Bunker 2012, 2013). Furthermore, guidelines and best practices for improving the social media communication of emergency agencies were developed in previous research (Freberg et al. 2013; Lin et al. 2016; Rice and Spence 2016; Ross et al. 2018). These guidelines were developed with the intent to optimise information diffusion in times of crisis and suggest, for example, to use social media as an active two-way communication channel and establish relationships to the public and other organisations (Lin et al. 2016).

## 2.2 Emergency Agencies and Social Media Analytics

In addition to employing it as another communication channel, social media offers sweeping advantages for emergency agencies. A study revealed that emergency management professionals perceive it as particularly important to analyse the large amount of available social media data as well to gain valuable insights during crisis situations and routine times (Fisher Liu et al. 2012). Emergency management professionals of the American Red Cross indicated that they make first attempts at analysing how the public speaks about them on social media and detecting crisis situations earlier by monitoring news accounts (Briones et al. 2011). According to them, they occasionally learn more quickly about a severe situation this way than if they would wait until they are informed by responsible authorities. A case study further reported about the State Control Centre Victoria which recently established a social media analyst position for improving situational awareness during crisis situations (Power and Kibell 2017). In this organisation, it was found that information retrieved by analysing user-generated content (e.g. photos from eyewitnesses) can be useful for planning the necessary resources to respond to an emergency situation early on. Related to this, researchers already developed technical solutions for predicting earthquakes based on the sentiment of tweets (Sakaki et al. 2010) or retrieving relevant information from social media content to improve situational awareness (Imran et al. 2018). However, the use of social media analytics in emergency agencies is still rather an exception due to a wide range of technological, organisational and environmental challenges such as a lack of guidelines, staff trained in data analysis and several legal issues related to the collection and analysis of social media data (Stieglitz, Mirbabaie, Fromm, et al. 2018). In sum, previous studies examined how social media is used during different crisis situations and how emergency agencies in particular use social media to communicate in times of crisis. Furthermore, there are prior studies focusing on the potential of social media analytics for early crisis detection or increasing situational awareness. However, the identification of further social media analytics use cases by examining the social media communication about emergency agencies represents a pivotal research gap.

## 3 Method and Results

### 3.1 Data Collection

In a first step, we used a self-developed Java tool which establishes a connection to the Twitter Search API to collect tweets and retweets including pre-defined keywords. The retrieved dataset included all tweets and retweets in the period from the 25$^{th}$ February 2018 to the 13$^{th}$ May 2018 in which the German emergency agency JUH was mentioned. For this purpose, the name of the emergency agency (johanniter), their slogan "for the love of life" (aus liebe zum leben) and abbreviations (johnnies) were defined as keywords. Synonyms, related spellings and hashtags were automatically considered by the tracking tool. At the same time, several keywords were excluded because they refer to the "Johanniter-Orden" which is an order of knights with a similar name. After a manual check of the dataset, 370 tweets were removed because it became apparent that the two keywords "johnnys" and "johnnies" were only included in tweets that had no connection to JUH. Overall, the revised dataset contained 1.085 tweets and 533 retweets.

## 3.2 Social Media Analytics

### 3.2.1 Frequency Analysis

We first conducted a frequency analysis to identify sudden increases in tweet activity which could indicate that an unusual discussion about JUH took place. To analyse the tweet activity over the course of the whole tracking period, we calculated the number of tweets and retweets per day. The visualisation tool Tableau was used to identify the highest peak and the resulting timeline can be seen in Figure 1.

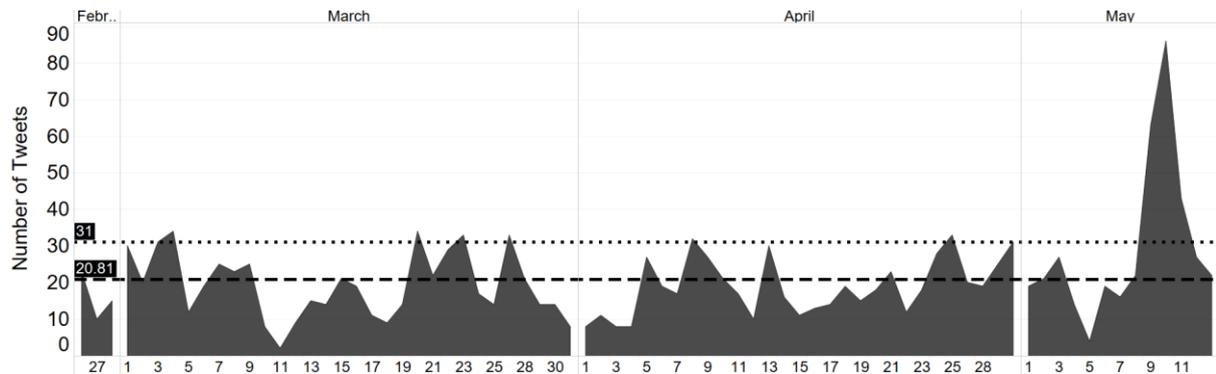

*Figure 1: Number of Tweets and Retweets per Day*

The average number of tweets and retweets mentioning the emergency agency JUH was 20.81 per day and the highest peak occurred from the 9th May 2018 to the 11th May 2018. On the 10th May 2018, the number of tweets mentioning JUH was 86 which is more than three times higher than the average tweet count.

### 3.2.2 Sentiment Analysis

In addition to the tweet activity, we also calculated the average sentiment of tweets and retweets per day over the course of the whole tracking period to examine the opinion of Twitter users about JUH. The sentiment within a tweet was extracted by using the opinion mining tool SentiStrength. The tool uses a dictionary-based sentiment detection and considers a set of additional linguistic rules such as negations, booster words and emoticons (Thelwall et al. 2010). In this research, the German sentiment dictionary was used. SentiStrength estimates the negative and positive sentiment strength of a tweet or retweet by reporting two scores (Thelwall et al. 2010). The overall sentiment score of a tweet is calculated by adding the two separate scores and can range from extremely negative (-4) to extremely positive (+4). Figure 2 displays the average sentiment of all tweets and retweets per day.

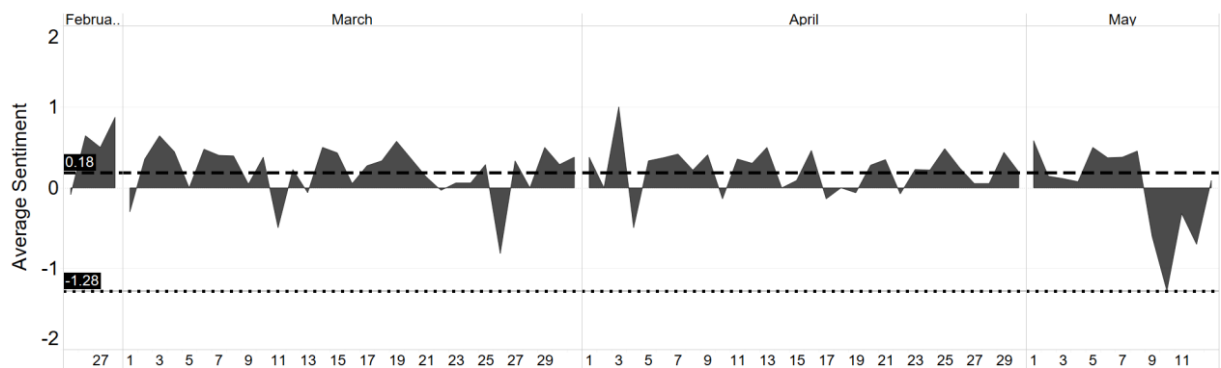

*Figure 2: Average Positive and Negative Sentiment per Day*

It can be seen that the communication about the emergency agency JUH was slightly positive most of the time as the average sentiment score per day was 0.18. The highest peak during the sentiment analysis could be identified on the 10th May 2018 with an average sentiment score of -1.28. This day constituted the highest peak during the frequency analysis as well.

### 3.2.3 Social Network Analysis

Furthermore, we conducted a social network analysis with the graph visualisation tool Gephi to identify influential Twitter users in the communication network around the emergency agency JUH. A retweet graph was created with 526 nodes representing Twitter users and 475 edges representing retweet activities. Then, we calculated the in-degree for each node to identify the ten most influential Twitter users. A high in-degree indicates that a node was more often retweeted than another node with a lower in-degree. Based on an existing categorisation of Stieglitz et al. (2017), the user roles of the ten nodes with the highest in-degree were identified by examining their user names and Twitter descriptions (see Table 1). The retweet network including the ten most influential nodes can be seen in Figure 3. Dyadic and triadic connections have been removed to enhance the visibility of the most influential nodes.

| User Name | Role | In-Degree |
| --- | --- | --- |
| JungeAlternative NRW | Political Organisation | 69 |
| Die Johanniter (JUH) | Emergency Agency | 64 |
| Johanniter Worldwide | Emergency Agency | 25 |
| AfD_Support | Individual (Political) | 20 |
| Hania Wiatrek | Individual (Political) | 15 |
| Oliver Janich | Individual (Political) | 13 |
| Polizei Mittelfranken | Emergency Agency | 12 |
| Bay. Rotes Kreuz | Emergency Agency | 12 |
| Politikversagen | Media Organisation (Political) | 9 |
| Stadt Dortmund | Government Organisation | 9 |

*Table 1. The Ten Most Influential Nodes and Their User Roles Sorted by In-Degree*

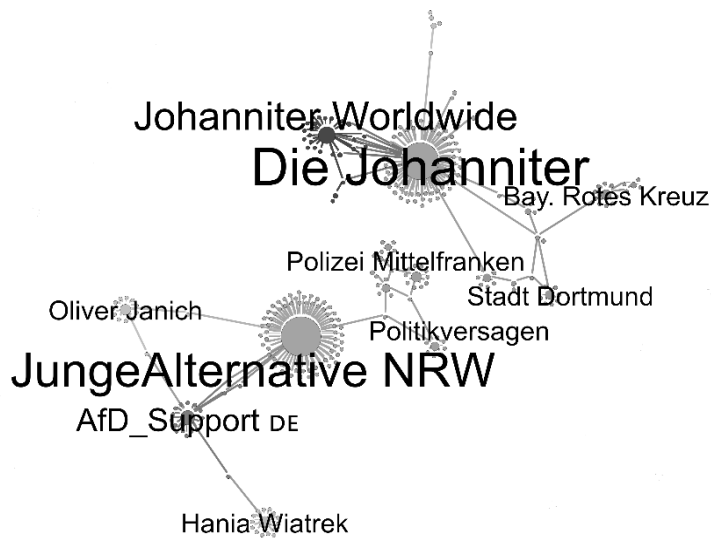

*Figure 3: Retweet Network Including the Ten Most Influential Nodes*

It can be seen that there are two main clusters in the network. The first one evolves around the German emergency agency JUH which only operates in Germany. The other influential users in this cluster can also be classified as emergency agencies. The account "Johanniter Worldwide" belongs to the emergency agency JUH but operates in foreign countries. The account "Bay. Rotes Kreuz" does not belong to the same organisation but has similar responsibilities like JUH. The remaining account "Stadt Dortmund" is the official account of a German city. The second cluster can be found around the German political organisation "JungeAlternative NRW" which is known for conservative and right-wing populist viewpoints. This account is closely connected to "AfD_Support" which is an unofficial Twitter account supporting the German right-wing party AfD. The cluster further includes two individuals and a media organisation both tweeting mainly political content as well as the account of a German police department.

### 3.2.4 Content Analysis

As a peak was identified from the 9th May 2018 to the 11th May 2018 during the frequency and sentiment analysis, we compared the content of the tweets created on these days with the content of the tweets created on the remaining days. We analysed the content by following the inductive category formation process of Mayring (2014). A category was built based on the content of the first tweet. The following tweets were either sorted into this category or it was decided to create a new category. As a result, the tweets were sorted into the following categories: information, job vacancy, rescue work, praise, criticism, call for action, attacks, warnings and crime. The tweets were manually coded by the research team and Cohen's kappa was calculated to assess the inter-coder reliability (Stemler 2001). Overall, the coding process was considered as reliable ($k = 0.79$). An example tweet for each category can be found in Table 2.

| Category | Example Tweet |
|---|---|
| Information | Attention can rescue lives: In our first aid tip, you learn how to spot a hypothermic person and how you can help. |
| Job Vacancy | Employee (m/f) for the department daily help. JUH GmbH, Location: Sankt August. |
| Rescue Work | #winterbus helps homeless people in Bremen (German city) |
| Praise | Dear voluntary helpers, thank you very much for your engagement at the #half-marathon in Berlin (German city)! |
| Criticism | Congratulations to Berlin JUH for the most stupid job commercial based on the motto "Come to us, here you have not even time to shit after work!" |
| Call for Action | Let us fight on together – for a better future! |
| Attacks | Attacks on rescue teams have become commonplace in #Merkel'sGermany. Most "incidents" are not made public because proceedings are terminated anyhow. #frustration #migration #cuddlyjustice #AFD (right-wing party) |
| Warnings | Caution! People who pretend to be JUH employees are collecting donations. |
| Crime | The wanted person apparently wears a patient bracelet with his name and the JUH logo. |

*Table 2. Example Tweets for Each Category*

The content of the tweets created during non-peak days can be described as follows: JUH was most often mentioned in informational tweets including first aid or health tips, information about staff changes, participation at events and the acquisition of new equipment. Furthermore, Twitter users often retweeted job advertisements of the emergency agency and attempts to recruit volunteers. Some tweets also included information about the rescue work and humanitarian projects of JUH. Moreover, the dataset contained tweets in which individual persons or other emergency agencies thanked the employees of JUH for their engagement in general or during specific events, emergencies and crisis situations. A smaller number of tweets included criticism related to the design of job advertisements, recruitment tactics of the agency, driving style of JUH employees, use of the siren at night, usability of the website and financing of the agency. Calls of JUH to participate in specific events such as first aid courses or to follow them on social media were also shared by a few users. Shortly before the highest peak, users started to retweet a news article about attacks on JUH paramedics during their work. The dataset further included a small number of warnings about persons who pretended to be JUH employees and dangers related to emergency situations or weather changes. The agency was also mentioned in a post of the police who was looking for a person wearing a bracelet with a JUH logo. Due to an incident on the 27th April 2018, the content of tweets created during the highest peak had changed considerably. In Figure 4, it can be seen which proportion of tweets were sorted into each category during peak and non-peak times.

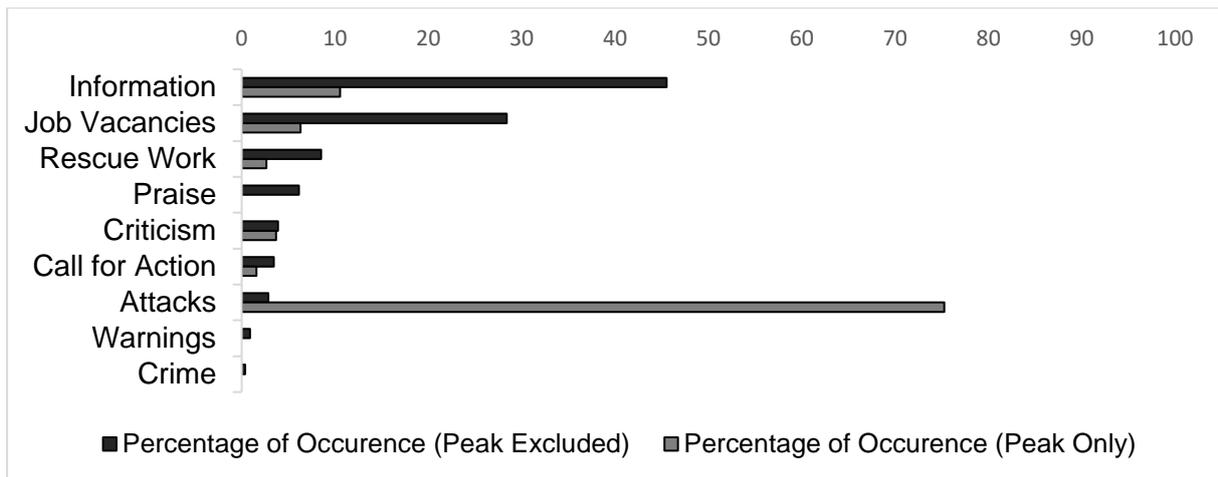

*Figure 4: Comparison of Content between Peak and Non-Peak Times*

As illustrated in Figure 4, the majority of tweets during peak time were sorted into the category "attacks". The German online tabloid newspaper "BILD.de" published an article about an interview with an employee of JUH. In this interview, the employee stated that he was attacked by drunken party guests while he wanted to help a 17-year old girl with shortness of breath. The article further included a quote of the employee stating that such attacks have become commonplace and a picture showing the arrest of a man of foreign origin. On Twitter, the German right-wing political organisation "JungeAlternative NRW" published a tweet with a link to this article and the hashtags #frustration #migration #cuddlyjustice and #AfD. In this tweet, the German chancellor is blamed because her migration policy is allegedly the reason for the frequent attacks. Similar tweets were posted by an individual person who claims to support the German right-wing party AfD and another individual person who regularly tweets about crimes committed by refugees in Germany. All three original tweets were frequently retweeted by members of the cluster around "JungeAlternative NRW" which was identified during the social network analysis. The emergency agency JUH was also mentioned in a few tweets criticising a job commercial in which an employee had to embark on a spontaneous rescue mission while sitting on a toilet. The remaining tweets were more neutral and contained information about the rescue work of the emergency agency and job vacancies. Figure 5 provides a comprehensive summary of the most important insights gained about the communication related to JUH through the applied social media analytics methods.

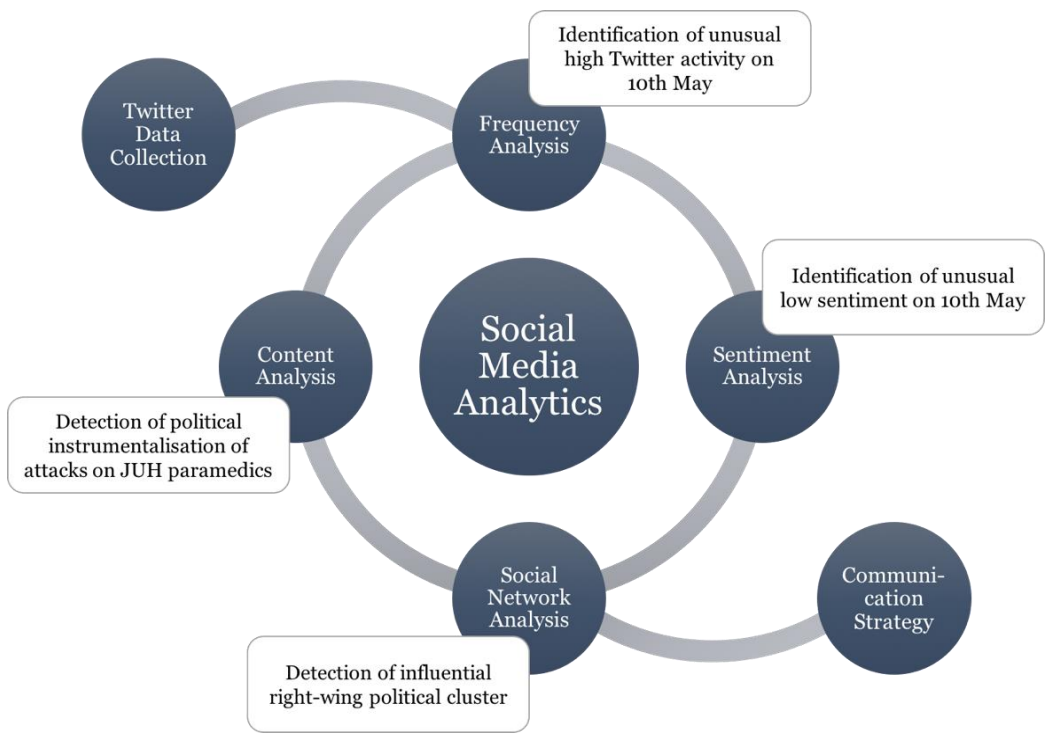

*Figure 5: Social Media Analytics Cycle for Emergency Agencies*

## 4   Discussion

The aim of this research was to examine which insights into the social media communication about an emergency agency could be gained by analysing available data on social media. The results reveal that social media analytics is well suited to identify important issues related to emergency agencies allowing them to adjust their communication efforts or implement organisational changes accordingly. In the following, we will discuss the benefits gained through the different applied methods of social media analytics in more detail.

The results show that peaks identified during a frequency and sentiment analysis could be a good indicator for unusual incidents related to an emergency agency. In this research, the tweet activity around JUH increased significantly after a news article about attacks on JUH paramedics was published. A sentiment analysis reveals that the emotionality of the Twitter communication about JUH became more negative at the same time. A peak during both types of analysis could be well suited to detect unusual incidents as the results revealed that the number of tweets mentioning JUH was usually rather low and the sentiment of the communication slightly positive. In previous studies, it was already shown that a sudden increase in tweets containing a specific hashtag in combination with a decrease in sentiment could be important feature to detect negative events such as earthquakes (Sakaki et al. 2010). For emergency agencies, it could therefore be relevant to examine the contents of such peaks more closely for being able to develop communication strategies for the identified incident.

The content analysis reveals that the communication during the frequency and sentiment peak differed significantly from the communication during non-peak times. By examining the content, it was found that the political instrumentalisation of attacks on JUH paramedics was the reason for the increase in tweet activity and the decrease in sentiment. While several news articles reported that drunk teenagers attacked JUH paramedics while trying to help a young woman, an article of a German tabloid newspaper about this incident included a picture showing how a man of foreign origin was arrested by the police. Neither media organisations nor the police officially stated that a man of foreign origin was responsible for the attacks. Nevertheless, the German right-wing political organisation "JungeAlternative NRW" shared the article on Twitter and blamed the migration policy of German chancellor Angela Merkel and the justice system in Germany for the attacks. The political instrumentalisation of the JUH-related incident is an example for an immigration threat narrative commonly disseminated by right-wing populist parties worldwide. A comparative study revealed that right-wing populist parties in the US, UK and Australia have in common that they use three types of immigration threat narratives (Hogan and Haltinner 2015). According to this study, right-wing populist parties share stories proposing that immigrants are either an economic, security or culture threat. The narrative created around the attacks on JUH paramedics can be classified as a security threat narrative which is based on the belief that "immigrants increase violence and property crime, bring diseases, and make the nation more vulnerable to terrorism" (Hogan and Haltinner 2015, p. 529). Our study contributes first insights on how social media platforms are used to create and spread such narratives. The study further revealed that news articles which present crimes in relation with people of foreign origin are used to spread right-wing populist narratives in the Twittersphere, which might be of particular relevance for police authorities. It was also shown that the highly emotional political tweets were retweeted much more often than tweets belonging to other categories suggesting that political instrumentalisation might reach more people than informational posts of emergency agencies. This is in line with a previous study revealing that political tweets with a negative sentiment are retweeted more often than neutral tweets (Stieglitz and Dang-Xuan 2013). Thus, emergency agencies should be aware of the possibility that incidents related to them could be politically instrumentalised to develop counteracting strategies for this case.

Another indicator for the political instrumentalisation of the attacks on JUH paramedics was the high influence of right-wing political Twitter users which became apparent during the social network analysis. The tweet of the right-wing political organisation "JungeAlternative NRW" was frequently retweeted by several other users with similar viewpoints. This way, the influence measured by in-degree of this organisation was higher than the influence of the agency JUH itself. This result is especially striking as this influence was only gained through tweets during the three-day peak while the tweets of JUH were retweeted throughout the whole period of data tracking. Employing social network analyses, thus, could be crucial for emergency agencies to identify influential Twitter users who do not belong to their own cluster. By relying solely on Twitter notifications, emergency agencies would only learn about users they are connected to or about users who are directly interacting with them or their tweets. This way, emergency agencies would probably not notice attempts at political instrumentalisation related to their organisation. This assumption is also supported by previous research which revealed that Twitter users are only seldom exposed to cross-ideological content as they are often part of clusters with a homogenous political orientation (Himelboim et al. 2013).

Analysing the social media communication about themselves provides emergency agencies not only with the opportunity to detect political instrumentalisation but also allows them to identify what users value about their work and what they are criticising. For example, some users mentioned that job advertisements of JUH were distasteful or not representing the current generation of elderly people properly. Emergency agencies could establish a dialogue with users who are criticising them to learn how they could improve their organisation in the future. However, the analysis also reveals that several individuals and emergency agencies were thankful for the engagement of the organisation's employees. Emergency agencies could try to foster their relationships to the individual persons in particular as these could be recruited as voluntary helpers. Strengthening their relationship to other emergency agencies might also be helpful for recruitment purposes as the content analysis revealed that these organisations frequently retweeted their job advertisements.

# 5 Conclusion

## 5.1 Summary

The aim of this research was to demonstrate the potential of social media analytics for improving social media communication of emergency agencies. A Twitter dataset was retrieved which included all tweets and retweets mentioning the German emergency agency JUH. Subsequently, we conducted a frequency, sentiment, social network and content analysis to examine by whom and in which context the particular emergency agency was mentioned on Twitter. It was found that a major peak occurred from the 9th May 2018 to the 11th May 2018 as a result of an interview about attacks on paramedics in a tabloid newspaper was used for political instrumentalisation by a cluster of right-wing political Twitter users. In the original article, it was not clear whether the paramedics were attacked by people of foreign origin. Nevertheless, the migration policy of the German chancellor Merkel was presented as reason for the attacks. These results indicate that emergency agencies should be careful how they communicate about incidents during their work which potentially involve people of foreign origin. The peak further included tweets of individuals criticising a job advertisement of the emergency agency for being distasteful. Based on the results, the emergency agency could improve future advertisements and develop strategies for avoiding political instrumentalisation of incidents related to them.

## 5.2 Limitations

In this research, we analysed the Twitter communication about the German emergency agency JUH. Therefore, the results of this paper should be interpreted carefully as they might not be transferable to other social media platforms as well as to different emergency agencies who fulfil other task or operate in another country. The choice of keywords represents a further limitation as it cannot be ensured that the defined list was exhaustive, and therefore some relevant tweets might not be included in the dataset. However, the keyword list was created in consultation with the emergency agency JUH to ensure that as many relevant keywords as possible were included. Another limitation is the use of a dictionary-based approach for analysing the sentiment of tweets as irony or sarcasm might be not detected.

## 5.3 Future Research

By adopting an explorative research approach, an example of political instrumentalisation of incidents related to an emergency agency was detected. It would be worth examining whether incidents or crisis situations to which emergency agencies respond are regularly used for political instrumentalisation. This would require monitoring social media communication about multiple emergency agencies and crisis situations over a longer time period. In this context, it would also be important to provide emergency agencies and other involved actors with strategies allowing them to counteract this phenomenon. Furthermore, it would be relevant to analyse the reactions of other users to assess the impact of political instrumentalisation on Twitter. Additional analyses such as a hashtag-co-occurrence network or a friendship-network could also provide a deeper understanding of the phenomenon. In addition, further studies involving different social media platforms (e.g. Facebook, blogs, reddit) and emergency agencies are required to gain more holistic insights. The results of the present study could also be used to train an algorithm in detecting political instrumentalisation as manual content analysis would not be applicable in larger datasets.